\documentclass[12pt]{article}
\usepackage{amsfonts}
\usepackage{epsfig}

\newcommand{\eq}{\begin{equation}} 
\newcommand{\eqx}{\end{equation}}
\newcommand{\eqn}{\begin{eqnarray}} 
\newcommand{\eqnx}{\end{eqnarray}}

\newcommand{\f}[2]{\frac{#1}{#2}}
\newcommand{\lra}{\longrightarrow}

\newcommand{\cor}[1]{\left\langle{#1}\right\rangle}

\renewcommand{\th}{\theta}
\newcommand{\sg}{\sigma}
\newcommand{\eps}{\epsilon}
\newcommand{\dl}{\delta}
\newcommand{\al}{\alpha}

\newcommand{\rr}[4]{#1, {\it #2 \/}{\bf #3} #4}
\newcommand{\qb}{\bar{q}}
\newcommand{\nn}{{\cal N}}

\newcommand{\pref}{\f{1}{2\pi \alpha'}}
\newcommand{\ttl}{\f{\tau^2 \th^2}{L^2}}

\newcommand{\tcl}{\f{\tau^2 \chi^2}{L^2}}
\newcommand{\Ttl}{\f{T^2 \th^2}{L^2}}
\newcommand{\Tcl}{\f{T^2 \chi^2}{L^2}}
\newcommand{\qqqq}{\quad\quad\quad}
\newcommand{\xpr}{x_\perp}
\newcommand{\rrr}{\mathbb{R}}

\begin{document}

\title{Minimal surfaces and Reggeization in the AdS/CFT correspondence}

\author{R.A. Janik$^{a,b}$  and R. Peschanski$^a$ \\ \\
$^a$Service de Physique Theorique  CEA-Saclay \\ F-91191
Gif-sur-Yvette Cedex, France\\
$^b$M.Smoluchowski Institute of Physics, Jagellonian University\\ Reymonta
4, 30-059 Cracow, Poland}

\maketitle

\abstract{We address the problem of computing scattering amplitudes related to 
the correlation function of two
Wilson lines and/or loops elongated along light-cone directions in
strongly coupled gauge theories.   
Using  the AdS/CFT correspondence in the classical approximation, the
amplitudes 
are shown to be related to minimal surfaces
generalizing the {\em helicoid} in various $AdS_5$ backgrounds.
Infra-red divergences appearing for Wilson lines
can be factorized out or can be cured by
considering the IR finite case of correlation  functions of two Wilson loops. 
In  non-conformal cases related to 
confining theories,  reggeized amplitudes with linear trajectories and
unit intercept are obtained and shown to
come from the approximately flat metrics near the horizon, which sets
the scale for the Regge slope. In the conformal case 
the absence of confinement leads to a different
solution. A transition between both regimes appears, in a confining
theory, when varying impact parameter.
}

\maketitle

\section{Introduction}

The theoretical calculation from ``first principles'' of high energy
scattering amplitudes in the so-called ``soft'' regime of QCD is  
among the oldest and yet unsolved problem of strong interaction physics. The 
main reason is that it requires a good understanding of 4-dimensional
gauge field theories at strong coupling which we do not possess till now. In
view of the  
recent developments of the AdS/CFT correspondence~\cite{ma98,ma99} it is thus 
natural to address this problem in the new setting proposed in this 
way. An exact  correspondence for QCD is not yet known, however useful
information can be obtained from known realizations for confining
theories.

We would like to discuss relevant physical properties of scattering
amplitudes at high energy expected from 
the S-Matrix theory of strong interactions~\cite{fr62}. In particular, 
Reggeization of scattering amplitudes is expected to occur, i.e. high-energy 
two-body amplitudes behaving as $A(s,t) = s^{\alpha(t)}\times (prefactors),$  
where $s,t$ are the well-known Mandelstam variables. $\alpha(t)$ is
the Regge  
trajectory corresponding to singularities of partial waves at
$j\!=\!\alpha(t)$ in the t-channel. Unitarity,
analyticity  and crossing  
relations  implied by the S-Matrix theory impose  constraints on
$\alpha(t).$ In particular the Froissart bound~\cite{fr61} implies
that $\alpha(t\!=\!0) \le 1$ and the prefactors of the amplitude are at
most like  $log^2 s$. Note that the Froissart bound assumes an  
underlying confining field theory, or at least a mass gap, since the
scale of the bound is fixed by the particle of smallest mass (e.g. the pion).

In~\cite{ja99}, we  considered  large impact parameter
and high energy scattering of colourless states   for $SU(N)$
supersymmetric gauge theories in the strong coupling, large $N$ limit
using the AdS/CFT correspondence. The gauge theory scattering
amplitude is linked with a correlation function of tilted Wilson
loops elongated along the light-cone
directions~\cite{Nacht,VV,kor,Nachtr}. 
In the AdS/CFT correspondence, these correlation functions  are related to
minimal surfaces in the $AdS_5$ geometry which have the Wilson 
loops as boundaries. The case considered in our previous paper
Ref.~\cite{ja99} 
involved disjoint minimal surfaces and thus the necessity of including
supergravity field exchanges  between the two corresponding string
worldsheets.  
The dominant contributions were identified and all correspond to real
phase shifts, i.e. purely elastic scattering. 
In particular, the contribution of the bulk
graviton gives an unexpected ``gravity-like'' $s^1$ behaviour of
the gauge theory phase shift in a specific range of energies and
(very) large impact parameters.

The main but stringent difficulty which limited the scope of
Ref.~\cite{ja99} was that the weak field approximation in supergravity 
was shown to be broken unless the impact parameter $L$ was
sufficiently large, namely  $\f{L}{a}\gg s^{2/7},$ where $a$ is the
transverse extension of the Wilson loop. 
If the above condition is not met, the produced gravitational field in
the dual AdS theory becomes strong, preventing perturbative
calculations to be done in this background.  

We will concentrate on a situation where the difficulty with
supergravity field exchanges does not arise, since there exists a
single connected minimal surface which gives the dominant contribution
to the scattering amplitude in the strong coupling regime, i.e. when
$\al' \!\to\! 0$. This will allow us to extend our study to small
impact parameters, where inelastic channels are expected to play an
important r\^{o}le.

In this approach  we will start by considering 
the correlation function of two Wilson {\em lines}
elongated along the two light-cone directions, 
a configuration which can be used for the description of 
high-energy quark-quark or quark-antiquark amplitudes in gauge 
theories~\cite{Nacht,VV,kor}. The r{\^o}le of the
quarks in the AdS/CFT correspondence will be played, as in
\cite{Wilson}, by the massive $W$ bosons arising from breaking
$U(N+1)\rightarrow U(N)\times U(1)$. The case of IR finite correlators
of Wilson loops will be dealt with in a second stage.

The plan of our paper is as follows: in section {\bf 2}, we will 
analyze the correlation function of Wilson lines leading to an
evaluation of $q\bar q$ and $q q$ scattering amplitudes at high  
energy. This will be done in the context of the black hole geometry in
AdS space~\cite{wi98} (static Wilson loops were first studied in
this background in \cite{rey,brand}), where one can use a
flat metric as a good approximation scheme near the horizon. We analyze the
factorizable structure of the IR divergences and isolate a cut-off
independent inelastic amplitude leading to reggeization.  
In section {\bf 3}, we consider the so-called ``conformal'' case of
the AdS/CFT correspondence for ${\cal N} =4 $ supersymmetric $SU(N)$
gauge theory, where the  
$AdS_5$ metric gives rise to a different minimal surface solution. 
The problem of the cancelation of the infra-red divergences is
analyzed by 
considering Wilson loop correlators in section {\bf 4}, leading to the
(approximate) derivation of scattering amplitudes between colourless
states, while the conclusions and open problems are pointed out in the final
section.

\section{Wilson lines and minimal surfaces in ``quasi-flat'' geometry}

\label{s.flat}

\begin{figure}[h]
\centerline{\epsfysize=6cm \epsfxsize=5cm \epsfbox{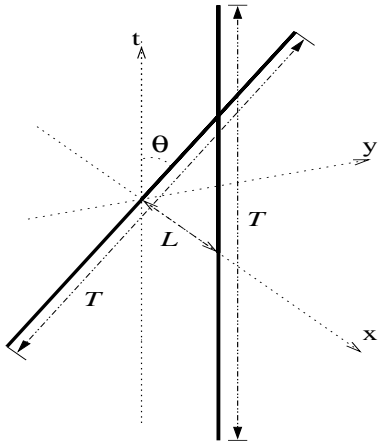}}
\caption{Geometry of the Wilson lines in euclidean space.}
\end{figure}

Let us start by defining an appropriate gauge theory observable for $q\bar q$  
scattering amplitudes $A(s,t)$.
It is convenient to pass from transverse momentum $t=-q^2$ to impact parameter 
space
\eq
\label{e.four}
\f{1}{s} A(s,t)=\f{i}{2\pi}\int d^2l \;e^{iq\cdot l}\;\tilde A(s,l)
\eqx
where $l$ is the 2-dimensional impact parameter (in the following we
will denote its modulus by $L$), and $\tilde A$ is the amplitude in
the impact parameter space. 

In the eikonal approximation the impact parameter space amplitude for
$q\qb$ scattering is
given by a correlation function of two Wilson lines~\cite{Nacht,VV,kor}  which
follow the classical straight line quark trajectories $W_1\lra
x_1^\mu=p_1^\mu\tau$ and $W_2\lra x_2^\mu=\xpr^\mu+p_2^\mu\tau$, with
$|\xpr|=L,$ see Fig.1. 
The IR cut-off will correspond to a fixed
temporal extent of the lines $-T<\tau<+T$.

The AdS/CFT correspondence gives a recipe \cite{Wilson,Gross} for
calculating this correlation function through
\eq
\label{e.minsur}
\cor{W_1W_2}\equiv \tilde A(s,l) = e^{-\f{1}{2\pi \al'} A_{minimal}}
\eqx
where $\cor{W_1W_2}$ is the   Wilson line 
correlator\footnote{The free propagation of the 
$q$ and $\qb$ states is not included in the correlator
$\cor{W_1W_2}$, which is thus implicitly normalized by
$1/\cor{W_1}\cor{W_2}$.}, $\al'=1/\sqrt{2g^2_{YM} N}$ in units of the
AdS radius, and 
$A_{minimal}$ is the area of the 
minimal surface in the appropriate background geometry
(e.g. $AdS_5\times S^5$ for
the conformal ${\cal N}=4$ SYM, an $AdS$ black hole \cite{wi98,rey,brand}
among other geometries \cite{estimates} for confining
theories) bounded by the Wilson 
line segments limited by the cut-off $T.$ A different approach to
discuss the minimal surface problem in the conformal $AdS_5$ was considered in
\cite{Zahed}, which concentrated on the elastic part of the
amplitude. 

Since the disjoint contour formed by the two Wilson line segments is
not closed, the procedure for finding a minimal surface is
ambiguous. We will adopt a prescription for finding the minimal surface
for infinitely long lines and then truncating it to a finite temporal
extent parameterized by the IR cutoff $T$. This implicitly consists of
forming a ``big'' Wilson loop closed at large temporal distance 
by curves drawn on the infinite minimal surface.

In turn, this procedure defines
the appropriate colour decomposition of the associated amplitude.
Using the well known colour decomposition $t^a_{ij} t^a_{kl}=-1/2N
\dl_{ij}\dl_{kl} +1/2 \dl_{il}\dl_{jk}$, we have
\eq
\tilde{A}(s,l)\equiv N
\left\{\tilde{A}_0(s,l)+\f{1}{2}\tilde{A}_{N^2\!-\!1}(s,l) \right\}
\eqx
where $\tilde{A}_0$ (resp. $\tilde{A}_{N^2-1}$) are the
amplitudes in the singlet (resp. adjoint) representations.

Using the same strategy  as in our first
paper~\cite{ja99}, we will perform the calculation with
euclidean signature for Wilson lines
in the boundary ${\mathbb R}^4$ forming a
relative angle $\theta$ in the longitudinal plane 
and then we will make an analytical continuation into Minkowski
space by rotating the euclidean time coordinate clockwise and the
angle anticlockwise (see \cite{Megg} in this context): 
\eqn
\label{e.anal}
\theta &\lra& -i\chi \sim -i\log \f{s}{m^2} \nonumber\\
T &\lra& iT \ .
\eqnx
Note that {\em a priori} there is an ambiguity in making the
analytical continuation depending on the precise choice of the path.
This phenomenon did not appear in the context of large impact
parameter near forward scattering discussed in \cite{ja99} since there,
the $\cor{WW}$ correlation function had only simple poles in the
complex $\th$ plane. In the case considered in this
paper the analyticity structure contains branch cuts
in the complex plane which have to be taken into account.

\subsubsection*{$AdS$ black hole solution and its flat space 
approximation}

In~\cite{wi98} a proposal was made that a 
confining gauge theory is dual to string theory in an $AdS$ black
hole (BH) background the relevant part of which can be written as
\eq
\label{e.bhmetric}
ds^2_{BH}=\f{16}{9}\f{1}{f(z)}\f{dz^2}{z^2} + \f{\eta_{\mu\nu}dx^\mu
dx^\nu}{z^2} + \ldots 
\eqx
where $f(z)=z^{2/3}(1-(z/R_0)^4)$ and $R_0$ is the position of 
the horizon\footnote{Compared to standard coordinates \cite{brand} we used
$U=z^{-4/3}$ and $U_T=R_0^{-4/3}$.}. 
Although it was later found that the $S^1$ KK states do not strictly
decouple in the interesting limits~\cite{de99}, we will use this background 
to
study the interplay between the confining nature of gauge theory and
its reggeization properties. Actually the qualitative 
arguments and
approximations should be generic for most confining backgrounds\footnote{
Two other geometries for (supersymmetric)
confining theories have been discussed recently \cite{ks,carlo}. 
They have the property that for
small $z$, i.e. close to the boundary, the geometry looks like $AdS_5\times
S^5$ (in \cite{carlo} up to logarithmic corrections related to asymptotic
freedom) giving a coulombic $q\qb$ potential. For large $z$ the
geometry is effectively flat. In all cases there is a scale, similar
to $R_0$ above, which marks a transition between the small $z$ and
large $z$ regimes.}, as already discussed in Ref.~\cite{estimates,so99}.

In order to calculate the scattering amplitude, we have to evaluate the
correlation function (\ref{e.minsur}). Therefore we put the
two tilted lines depicted in Fig.~1 on the boundary at $z=0$. Next we
have to find the minimal surface in the appropriate geometry which has
the two lines as its boundaries. The relative angle (tilt) in the
$t$-$y$ plane and the separation in the transverse direction $x$
(impact parameter) therefore define the boundary conditions for the
geodesic equations for the string.

As is well known for the Plateau problem of minimal surfaces
\cite{fomenko} the
boundary conditions determine the solutions. 
Although an exact solution for the minimal
surface spanned by the tilted Wilson lines is unknown for the metric
(\ref{e.bhmetric}), the properties of the black hole (BH) geometry allow
for quite a good approximation scheme.

Two salient features of the metric (\ref{e.bhmetric}) are (i) the standard AdS
prefactor $1/z^2$ close to the boundary ($z=0$), (ii) the existence of
a horizon which limits from {\em above} the values of $z$. A
consequence of (i) is that it is most efficient for a minimal surface
to perform the ``twisting'' between the two Wilson lines as far away
from the boundary as possible. Property (ii) effectively induces this
twisting to occur near the horizon as we shall show below.

The appropriate minimal surface in the BH geometry
will look as follows. Due to property (i), the minimal 
surface between well separated lines  rises
``vertically'' in the $z$ 
direction up to the horizon without sizable motion in the other $\rrr^4$
coordinates (see a schematic representation in Fig.~2). 
The metric at the horizon is effectively flat 
\eq
\label{e.methor}
ds^2_{\sim horizon}\sim \f{1}{R_0^2}\ (\eta_{\mu\nu}dx^\mu dx^\nu) \ ,
\eqx
and the motion in the $z$ direction is ``frozen out''.
Now near the horizon, following property (ii), the minimal
surface performs the ``twisting'' (not displayed in Fig. 2) 
corresponding to the tilt angle
$\th$ between the initial Wilson lines.
At this stage we thus have to find a minimal surface between
the lines at an angle $\theta$ in the flat space metric
(\ref{e.methor}). Finally the surface falls off again vertically
towards the boundary. The area of the ``vertical'' pieces is removed by
the standard subtractions \cite{Wilson}, so the resulting area which enters the
formula for the amplitude (\ref{e.minsur}) may be approximated by the area
of the ``flat space'' piece.

\begin{figure}
\centerline{\epsfysize=5cm  \epsfbox{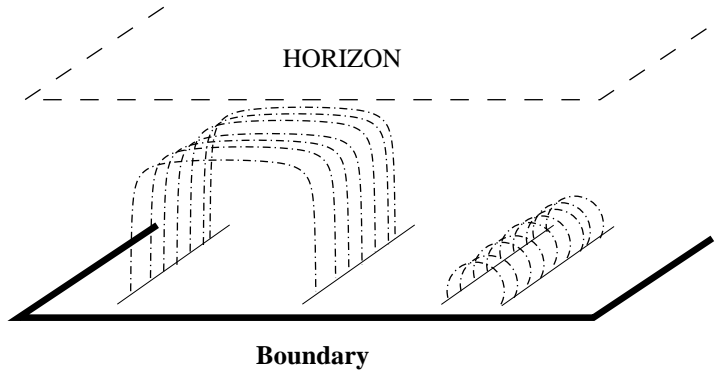}}
\caption{The minimal surface in the black hole geometry. The Wilson
lines are drawn here with vanishing angle of tilt $\th=0$.}
\end{figure}

We will now substantiate this intuitive picture with a more quantitative
study of the geodesic equation for the string. Let us determine under
what conditions the minimal surface 
spanned by the two tilted Wilson lines is indeed predominantly flat and
concentrated near the horizon following the general line of discussion
of \cite{estimates}.

The minimal surface equations follow from the Nambu-Goto action.
\eq
\label{e.action}
S=\f{1}{2\pi \al'}\int_{-T}^Td\tau
\int_{-l(\tau)/2}^{l(\tau)/2}\sqrt{\det h_{ab}},
\eqx
where the induced metric on the worldsheet is
\eq
\label{e.indmetric}
h_{ab} \equiv G_{ij}\f{\partial X^i(\sg,\tau)}{\partial v^a}
\f{\partial X^j(\sg,\tau)}{\partial v^b}.
\eqx
$X^i$ stands for general coordinates $(z,x^\mu)$ in
(\ref{e.bhmetric}), $G_{ij}$ is the background metric, $v^0\equiv \sg$,
$v^1\equiv \tau$, and $l(\tau)=\sqrt{L^2+\th^2 \tau^2}$ is the euclidean 
distance between points on the two Wilson lines with the same value of
the time coordinate $\tau$. 

As a first remark we note that using the background metric
(\ref{e.bhmetric}) the terms in the induced metric $h_{\sg\sg}$
corresponding to the twisting are of the form
\eq
\f{1}{z^2}\left[\left( \f{\partial y}{\partial \sg}\right)^2 + \left
( \f{\partial t}{\partial \sg}\right)^2 \right].
\eqx
Hence, near the boundary ($z \to 0$), the minimization will not
change noticeably the twist angle. Thus the boundary conditions are
``frozen'' and transported to the vicinity of the horizon. 

For further discussion we shall make an
approximation (similar 
to \cite{Zahed}) of neglecting explicit $\tau$ dependence in the
Euler-Lagrange equations following from (\ref{e.action}) and leaving
it only in the implicit dependence on the boundary conditions through
$l(\tau)$. Within this approximation the estimate of \cite{estimates},
made for the case of the static $q\qb$ potential may be directly
applied to our problem.

In reference \cite{estimates} a distance $d$ is defined, for all
metrics giving confinement, which
measures the transverse distance (on the boundary) over which the
string worldsheet significantly deviates  from being flat. In all cases
the ratio $d/l(\tau) \lra 0$ when $l(\tau)\to \infty$..
Depending on the confining metric considered, $d$ behaves as a
logarithm or a power of $l(\tau)$ smaller than one. 
It is interesting to note that the condition $d/l(\tau) \ll 1$ leads to
a lower bound on the impact parameter $L$ since the above condition is
most restrictive for the smallest value of $l(\tau)$, which is equal to
$L$.

The precise dependence of $d(l(\tau))$ on the horizon scale $R_0$
depends on the metric considered.
For instance for the metric (\ref{e.bhmetric}) rescaling
arguments lead to a dependence $d(l(\tau))\sim R_0^{2/3} \log l(\tau)$.
Therefore as long as the impact parameter is large
with respect to $R_0$ the approximations considered in this section
should be valid.  

However, it may of course happen that the impact parameter distance between the
two Wilson  
lines becomes much smaller than $R_0.$ In this case (see Fig.~2)
the minimal surface problem becomes less affected by the black hole
geometry (or the large $z$ behaviour of the different metrics
\cite{ks,carlo}) and will just probe the small $z$ region of the
geometry.
 
The precise behaviour at these shorter distances will depend on the
type of gauge theory and, in particular, on the small $z$ limit of the
appropriate metric. In 
this paper we will consider the generic case (from the 4D (S)YM point
of view) when this limit resembles the original $AdS_5\times S^5$
geometry~\cite{ma98}. We will consider this conformal (non confining)
regime in detail in a further section. We note that the
same behaviour can be equivalently obtained through rescaling, by
keeping the impact parameter fixed and putting the scale $R_0 \to  \infty$.

Let us concentrate in the following on the case when the
impact parameter is larger than the scale $R_0$.
To summarize the discussion, the string is then to
a large degree concentrated in the region near the horizon
(\ref{e.methor}) with the boundary conditions essentially transported
from $z=0$. 
We are thus led first to calculate the area of the minimal surface
bounded by the tilted
lines in the {\em flat} geometry (\ref{e.methor}) {\em at} the
horizon. We will first perform the calculation in euclidean signature
and then perform the analytical continuation (\ref{e.anal}).

\subsubsection*{Helicoid geometry}

The basic building block of our construction is a minimal surface
spanned by two straight line segments of length $2T,$ corresponding to the two 
Wilson lines
 separated by a
distance $L$ in the ``transverse'' direction $x$ and with a relative angle
$\theta$ in the ``longitudinal'' plane:
\eq
L_1:\tau \lra (\tau,0,0,0) \qqqq L_2:\tau \lra (\tau \cos\th,\tau \sin 
\th,0,L).
\eqx
It is well-known that in the flat ${\mathbb R}^4$ geometry the minimal surface 
with infinite boudaries $\tau=-\infty\ldots\! +\!\infty$ is a 
helicoid. We will also be interested by the ``truncated'' 
helicoid\footnote{For finite cut-off $T$ in flat space, the truncated
helicoid obviously remains a solution if one adds the boundary helices
at $\tau=-T, T$ 
as new boundaries. Note, however, that 
with these boundaries the helicoid may be an unstable~\cite{bo99}
minimum for a too large value of the cut-off. We will not consider
this problem in the present paper.} where $\tau=-T\ldots T$. 

Let us recall the minimal surface solution in flat space. 
The helicoid is the only regulated 
(spanned by straight lines) minimal surface. The truncated 
helicoid solution may be parametrized by
\eqn
t&=&\tau \cos \f{\th \sg}{L}\nonumber\\
y&=&\tau \sin \f{\th \sg}{L}\nonumber\\
x&=&\sg
\eqnx
where $\tau=-T\ldots T$ and $\sg=0 \ldots L$ and $\th$ is the total
twisting angle.

Its area is given by the formula
\eqn
\label{e.ahelic}
Area \equiv S(T)=\int_{0}^{L}d\sigma \int_{-T}^{T}d\tau \sqrt{1+\ttl}=
\nonumber\\
=LT\sqrt{1+\Ttl} +\f{L^2}{2\th} \log
\f{\sqrt{1+\Ttl}+\th \f{T}{L}}{\sqrt{1+\Ttl}-\th \f{T}{L}} \ .
\eqnx
Let us now perform the analytical continuation (\ref{e.anal}), which
links euclidean correlation functions in gauge theories with
minkowskian ones directly related to scattering amplitudes.
A naive
continuation of the area formula (\ref{e.ahelic}) leads to a pure
phase factor in (\ref{e.minsur}):
\eq
\label{e.ahelicmin}
\exp \left\{\f{\sqrt{2g^2_{YM}N}}{2\pi R_0^2}  i\left[ LT\sqrt{1+\Tcl}
+\f{L^2}{2\chi} \log \f{\sqrt{1+\Tcl}+\chi
\f{T}{L}}{\sqrt{1+\Tcl}-\chi \f{T}{L}} \right] \right\} \ , 
\eqx
where $1/2\pi\al'$ in (\ref{e.minsur}) has been replaced by the
factor
$\sqrt{2g^2_{YM}N}/(2\pi R_0^2)$ coming from the flat metric
(\ref{e.methor}).

However the
analytic structure of the euclidean area (\ref{e.ahelic}) involves 
cuts in the complex $T$, $\th$ planes and thus leads to an ambiguity
coming from the branch cut of the 
logarithm. In fact when performing the analytical continuation we have
to specify the Riemann sheet of the logarithm (i.e. $log \to log+2\pi
i n$). This leads to an additional real multiplicative factor in
(\ref{e.minsur}): 
\eq
\label{e.ai}
\exp\left\{ - n \f{\sqrt{2g_{YM}^2N}}{\chi} \f{L^2}{2R_0^2} \right\}\ ,
\eqx
the form of which is {\em uniquely} fixed by the euclidean expression
(\ref{e.ahelic}) up to a choice of the integer $n$. Within the
classical approximation which we have been using it is not possible
to determine the value of $n$. On a more physical ground, in
section~{\bf 4}, we will relate the analogue of the label $n$ which appears in
the calculation of Wilson loop correlators with
multivalued saddlepoint minima of a minimization equation and thus to
different classical solutions.
The determination of the relative weights of the
various contributions goes beyond the classical approximation used
throughout this paper\footnote{We also note the close similarity of
the $n$, $L$ and $\chi$ dependence in (\ref{e.ai}) with an analogous
factor $\exp(-nL^2/\pi\chi)$ in the {\em imaginary} part of the D
brane scattering amplitude \cite{ba94} where $n$ labels the poles of
the appropriate string partition function between the branes.}.

As can be seen the contribution (\ref{e.ai}) is cut-off {\em independent}.

Another useful way of deriving the above factor (\ref{e.ai}) can
be directly obtained from the integral leading to (\ref{e.ahelic}).
This method can be generalized to  more complicated background geometries, for
instance to the conformal case, which we will consider later, for which
we lack an exact expression of the form (\ref{e.ahelic}).

Let us perform only the first part of the analytical continuation
(\ref{e.anal}) 
$\th \lra -i\chi$ but otherwise remain with the time variable $T$ in Euclidean
space. This procedure yields the expression:
\eq
\label{e.semi}
\int_{-\f{L}{\chi}}^{\f{L}{\chi}} d\tau \int_0^L d\sg
\sqrt{1-\tcl}=\f{\pi L^2}{2\chi}\ . 
\eqx
We see that the imaginary part may be obtained by integrating 
($n$ times) around
the branch cut of the square root. A convenient reinterpretation of
the above formula follows from
performing the change of variables $\sg\lra \sg'=\sg
\sqrt{1-\tcl}$. Then we get
\eq
\label{e.semiel}
2in\int_{-\!\f{L}{\chi}}^{+\!\f{L}{\chi}} d\tau
\int_0^{\sqrt{1-\tcl}L} d\sg'=ni\pi\f{L^2}{\chi}
\eqx
which is effectively twice ($\times in$) the area of a minimal surface 
bounded by a `half-elipse' of radii $L$ and $L/\chi.$ This $T$-independent
imaginary part is unaffected by the second part of the analytical
continuation (\ref{e.anal}), and leads directly to the factor
(\ref{e.ai}).

\subsubsection*{Reggeization in quark-(anti)quark scattering}

Our result for the Wilson line correlation function for the $AdS$ BH
geometry gives rise to the following contributions
\eqn
\label{e.ampfull}
\tilde{A_n}\!\!\!&\!=\!&\!\!\!\exp\left\{ \f{\sqrt{2g^2_{YM}N}}{2\pi
R_0^2}  i\left[ LT\sqrt{1\!+\!\Tcl} 
\!+\!\f{L^2}{2\chi} \log \f{\sqrt{1+\Tcl}\!+\!\chi
\f{T}{L}}{\sqrt{1\!+\!\Tcl}\!-\!\chi \f{T}{L}} \right]\right\}
\nonumber\\ 
&& \times \ \exp\left\{ - n \f{\sqrt{2g_{YM}^2N}}{\chi} \f{L^2}{2R_0^2}
\right\} \ .
\eqnx
There is a divergent phase in the above amplitude
when the temporal length of the lines $T$ goes to infinity. We 
interpret this divergence as reflecting  the expected
IR divergence of the $q\!-\!\qb$ scattering
amplitude~\cite{VV,kor}. A consistent way to eliminate this cut-off
dependence is to consider an
IR finite physical quantity like scattering  of two  $q\qb$ pairs (see 
section {\bf 4}). In the present case of Wilson lines,
the specific factorized form of
(\ref{e.ampfull}) allows for a determination of an IR finite
contribution, which can be interpreted as an effect of inelastic
channels on the Wilson line correlator. 

It is known since a very long time that the superposition of long
range and short range potentials in the Schroedinger equation 
leads to a factorization formula for 
the relevant S matrix elements for each partial wave~\cite{Messiah}. 
For instance in nuclear
physics, the superposition of long range coulombic and
short range interactions leads to a factorization into the
elastic coulombic S matrix element and a short range
amplitude modified by the long-range background.
The elastic S matrix may be treated as a redefinition of
the asymptotic initial and final states. The amplitude reads
\eq
\label{e.tdef}
A(l,s)=e^{2i\dl(l,s)} \cdot T(l,s) 
\eqx
where $\dl$ is the {\em real} phase shift due to the
elastic long range interactions and $T$ is the short range part of the
amplitude.
For instance in the QED result for electron scattering
\cite{qedeik}, the real phase shift exhibits a divergence which can be
written as   
\eq
\label{e.qed}
e^{2i\dl(l,s)} \propto \exp\left\{i\f{e^2}{4\pi}\coth \chi
\log\left(\f{L^2}{4T^2}\right)\right\} \ ,
\eqx
where $1/T$ has been  substituted for a fictitious photon mass (IR
regulator).

In hadronic interaction physics \cite{bvh}, a similar factorization
appears for the S matrix elements for 2-body channels in terms of an
elastic contribution and 
an amplitude $T(l,s)$ which, by unitarity of the S
matrix, arises from the
contribution of many inelastic channels to the 2-body S matrix \cite{vh}.
In this context the amplitude (\ref{e.tdef}) can be related to the
inelasticity (overlap matrix) in the scattering namely
\eq
\label{e.bhv}
T(l,s)=\f{1-\sqrt{1-2f(l,s)}}{2}
\eqx
where the overlap matrix elements $f(l,s)$ are defined from the 2-body
S matrix contribution to unitarity $|S(l,s)|^2 \equiv 1-2f(l,s)$.

We are led to interpret our resulting amplitude (\ref{e.ampfull})
in the same way. The factor (\ref{e.ahelicmin}) can be treated as
redefining the initial and final $q\qb$ states due to long range
interactions\footnote{It is clear that there remains a freedom in
attributing a finite real phase shift either to the redefinition of
the states or to the interaction. Here we adopt the convention that
$T(l,s)$ is purely real and thus contains information only on the
inelasticity.}.
Naturally this phase is IR
divergent. The analogous inelastic contribution $T_n(l,s)$ is
obtained to be the cut-off independent factors (\ref{e.ai}).
Note that this physical interpretation requires the integer $n$ to be
positive. We will return to the discussion of the $n$ dependence in a
further section.

Let us discuss both factors of the amplitude (\ref{e.ampfull}). The
contribution of the real phase shifts behaves in the large $T$ limit like
\eq
\exp\left\{\f{\sqrt{2g^2_{YM}N}}{2\pi R_0^2} i\left(T^2 \chi
+\f{L^2}{\chi} \log\left(\f{2\sqrt{e}\chi T}{L}\right) \right)+O(1/T^2)
\right\}
\eqx
The appearance of the IR divergent $T^2$ and $L^2 \log T$ terms in the phase
shift can be linked with the linear confining potential of the theory. 

The effect of the confining potential is expected to
generate inelastic channels through the phenomenon of string
breaking and/or closed string emission. Within the above framework, where we
select initial and final $q\qb$ 
states, this contribution is expected to appear as an inelastic real
factor in the amplitude, while the phase factor diverges with $T\lra
\infty$. 

The inelastic $q\qb$ interaction amplitude at level $n$ is
\eq
\label{e.ampim}
T_n(l,s)=\exp\left\{ - n \f{\sqrt{2g_{YM}^2N}}{\chi}
\f{L^2}{2R_0^2} \right\}
\eqx
where the initial and final states are both $q\qb$.
It can be easily fourier transformed into transverse momentum space giving
\eq
\label{e.20}
T_n(s,t)=\f{iR_0^2\ln s}{n \sqrt{2g^2_{YM}N}}\ s^{ 1+\f{R_0^2}{2n
\sqrt{2g^2_{YM}N}}t} \ . 
\eqx
This contribution is thus reggeized with a linear Regge trajectory
with unit
intercept and the slope given by the string tension related to
the horizon distance $R_0^2$. 

It is worthwhile to consider what changes in the preceeding
discussion if we go from $q$-$\bar{q}$ scattering to $q$-$q$
scattering. In geometric terms, this corresponds to changing the
orientation of one of 
the lines, and since the string worldsheet spanned on the Wilson lines
is oriented, the twisting  angle of the helicoid changes as 
\eq
\label{e.twist}
\th \lra \th-\pi
\eqx
Upon analytical continuation this means that $\chi\sim \log s$
changes to $\chi-i\pi\sim \log se^{-i\pi}$, as required by
crossing properties, which are seen to have here a very simple geometric
interpretation.  
We note that in the asymptotically high energy limit $\log s \gg 1$,
one obtains the same factors (\ref{e.ampfull}) for both $qq$ and
$q\qb$ channels. Keeping the next to leading correction corresponding to  
$\log s \lra \log se^{-i\pi}$ preserves the crossing
relations between those channels.

Finally let us compare our result with the general structure of Wilson
line correlators at weak gauge coupling.
Indeed, the large $T$ dependence of the $q\qb$ amplitudes we discuss
reflects IR 
divergences which appear already in perturbative (weak coupling)
calculations of the same quantities.

For instance in the case of QED the whole dynamics is contained in the
infinite phase factor (\ref{e.qed})  and the divergence is logarithmic.  

The (renormalon improved) 1-loop QCD result \cite{kor,korch} for
$q\qb$ scattering is 
\eq
\label{e.qcdnonp}
\exp\left\{-\f{1}{\chi}
\f{\al_s}{\pi}\log\left(\f{T}{L}\right)-\f{\rho}{\pi} \Lambda^2
\f{L^2}{\chi} \right\}
\eqx
where $\rho$ is an undetermined nonperturbative parameter.
We note the compatibility between the
nonperturbative cut-off independent piece in (\ref{e.qcdnonp}) and
an analogous term in our result (\ref{e.ampim}). Our nonperturbative
result gives a hint on the scale and coupling dependence.

If the intercept one common to all contributions $T_n(s,t)$ (see
formula (\ref{e.20})) is not spoiled by the different weights
corresponding to fluctuations of the worldsheet around classical
solutions, it would be a candidate for the intercept one trajectories
(the so-called pomeron and odderon)
which are expected to emerge from a confining strongly
interacting gauge theory.

\section{Conformal case}
\label{s.conf}

The flat metric approximation  which we have used to derive the 
resulting area
(\ref{e.ahelic}) assumed that the impact parameter $L$ is
sufficiently large with respect to the scale set by the horizon radius
$R_0$ (or a similar scale in the backgrounds \cite{ks,carlo}
interpolating between a confining geometry at large $z$ and
approximately $AdS_5\times S^5$ near the boundary $z=0$).
In this regime the dominant contribution to the amplitudes
came from the part of the string worldsheet stretched near the horizon.

If we go to smaller impact parameters $L<R_0$ (and also for
$T<R_0$), the minimal 
surface would only penetrate into a limited region near the boundary
$z=0$, see Figure 2. 
In the scenarios which behave better at
short distances  than the original BH proposal, the metric becomes
closer and closer to the conformal $AdS_5$ case. We note that the
$AdS_5\times S^5$ setting is directly related to scattering in the
$\nn=4$ SYM. This different geometry leads to a qualitatively new
behaviour which we now analyze.

\subsubsection*{The conformal $AdS_5$ case}

In the case of $\nn=4$ SYM corresponding to the $AdS_5\times S^5$ background we
do not know yet the exact generalization of the helicoid, and some
approximation scheme is needed. As in the previous case, we 
will  concentrate on extracting the inelastic contribution which
appears also here to be  independent of the IR temporal cut-off $T$. We
use the method outlined in 
section {\bf 2} leading to formulae (\ref{e.semi})-(\ref{e.semiel}). 

Within a variational approximation approach,
we will look for a minimal solution in a restricted set of surfaces
(``generalized helicoids'') parameterized by
\eqn
\label{e.genhel}
t&=&\tau \cos \f{\th \sg}{L}\\
y&=&\tau \sin \f{\th \sg}{L}\\
x&=&\sg\\
z&=&z(\sg,\tau)\ .
\eqnx
Evaluation of the induced metric gives rise to the following area
functional:
\eq
\int_{-T}^{T} d\tau \int_0^L d\sg \f{1}{z^2} \sqrt
{ \left(1+\ttl\right) (1+z_\tau^2) +z_\sg^2 } \ .
\eqx
We perform a change of variables $\sg \lra \sg'=\sg
\sqrt{1+\ttl}$ which yields
\eq
\label{e.acft}
\pref \int_{-T}^{T} d\tau \int_0^{L\sqrt{1+\ttl}} d\sg'
\f{1}{z^2}\sqrt{1+z_\tau^2+z_{\sg'}^2} \ . 
\eqx

As in the previous section the cut-off independent part is obtained
from the branch cut structure of the area functional (\ref{e.acft}).
The analytic continuation $\th \lra
-i\chi$ changes the boundary conditions for the minimal surface
to be a half elipse of width $L/\chi$ and height $L$ (the upper
integration limit in (\ref{e.acft}) then becomes $L\sqrt{1-\tau^2
\chi^2/L^2}$). Due to conformal
invariance we know that the minimal area has the following form:
\eq
\label{e.adecomp}
A_{minimal}=f(L/\eps,\chi)+g(\chi)
\eqx
where $\eps$ is the ``$5^{th}$'' AdS coordinate where we put the 
D3 brane probe. $\eps$
translates directly into the mass of the $W$ bosons which play here the
role of quarks.
We do not expect higher poles in $\eps$ than first order, which
are in the standard way subtracted out \cite{Wilson,neww}, so we have at most a
logarithmic behaviour in $L/\eps$. 

It is possible to obtain an approximate result in the high energy
$\chi\! \lra\! \infty$ limit from known properties of  
Wilson loop expectation values \cite{Wilson,neww}.
The half-elipse has two cusps each with an angle $\pi/2$, whose
contribution to $A_{minimal}$ can be obtained from the results of
\cite{neww}. This leads to the following logarithmic terms:
\eq
-2\f{1}{2\pi}F(\pi/2) \cdot \log\f{L}{\eps \chi}
\eqx 
where $F(\Omega)$ is a complicated function calculated in \cite{neww}
($F(\pi/2)\sim 0.3 \pi$). 
The $\eps$ independent term $g(\chi)$ in (\ref{e.adecomp}) 
can be approximated by
noting that at high energies the half-elipse is very much elongated
and looks like parallel lines of
length $L$, roughly $2L/\chi$ apart. An approximate evaluation is then
given by integrating the coulombic potential \cite{Wilson}:
\eq
\label{e.intads}
-c\cdot\int_0^L \f{d\sg'}{\f{2}{\chi}
\sqrt{L^2-\sg'^2}}=-c\f{\pi}{4}\cdot \chi
\eqx
where $c=8\pi^3/\Gamma^4(1/4)$ is the coefficient in front of the
(screened) coulombic potential.
So we get
\eq
\label{e.lcft}
T_n(l,s)\sim\left(\f{L}{\eps\log s}\right)^{n\f{F(\pi/2)}{\pi} \ 
\f{\sqrt{2g^2_{YM}N}}{2\pi}} 
\  s^{n\f{ 2\pi^4}{\Gamma(1/4)^4} \  \f{\sqrt{2g^2_{YM}N}}{2\pi}} \ .
\eqx
Here, as in the case of the confining theory, the values of $n$ and
the weights of the different components $T_n(l,s)$ are not specified.

Let us comment on the behaviour of the various components. 
In all cases we  obtain a factorized energy  behaviour with no moving Regge
trajectories. We note that for $n$ positive a similar energy
dependence (i.e. with 
intercept greater than 1 and a (nearly) flat Regge trajectory) is obtained by
resumming the leading $\log s$ terms in the perturbative expansion at
weak coupling~\cite{BFKL} for the singlet exchange amplitude. 
In the conformal case, there
remains a non-perturbative screening effect (already present for the
static $q\qb$ potential~\cite{Wilson}) which appears as the change
$g^2_{YM}N\! \to\! \sqrt{g^2_{YM}N}$ in the exponent of $s$. 

Considering the impact parameter dependence and its Fourier transform
to momentum space, in the window of 
convergence (the exponent of $L$ in (\ref{e.lcft}) between $-2$ and
$-3/2$), we get 
\eq
\label{e.adsregge}
T_n(s,t)\sim i s^{1+n\f{ 2\pi^4}{\Gamma(1/4)^4} \cdot
\f{\sqrt{2g^2_{YM}N}}{2\pi}} 
\left(\f{1}{t}\right)^{1+n\f{F(\pi/2)}{2\pi} \ 
\f{\sqrt{2g^2_{YM}N}}{2\pi}}
\eqx
Otherwise one observes either an UV divergence 
(for exponent values less than $-2$) or an IR one (for values larger
than $-3/2$). 
For positive values of $n$ the IR divergence requires a careful
treatment which is beyond the scope of this paper. Note that, in the
case of ${\cal N}=4$ SYM, it can lead to infra-red divergent
pieces also in the inelastic amplitude, as is the case already in the
perturbative limit \cite{kor,bern}.

\subsubsection*{Conformal/non-conformal transition}

As  already mentioned, the
result (\ref{e.adsregge}) obtained for the pure $AdS_5\times S^5$ case should
give the dominant behaviour also for the confining theory for impact
parameters small with respect to the horizon scale\footnote{We need
also a sufficiently small $T$ parameter.} $R_0$ (or more generally
an analogous transition scale in the geometries \cite{ks,carlo}). 
Indeed this $R_0$ provides a
natural value of the impact parameter cut-off $L_0$. 
Thus even in the confining
theory, when the impact parameter is decreased and gets smaller than
$R_0$, we expect a transition from the set of components
(\ref{e.ampim}) to the results (\ref{e.lcft}), as long as the relevant
geometry for small $z$ is similar to  $AdS_5\times S^5$.

This process can be observed by noting that both (\ref{e.ampim}) and
(\ref{e.lcft}) were derived from a minimal surface spanned on a
semielipse. The result for impact parameters $L\geq O(R_0)$ was obtained by
using the area law for Wilson loops, while the  
conformal case (corresponding here to $L\ll R_0$) 
used an approximation using coulombic potential. The
solution of the appropriate minimal surface problem in the full
geometry would lead to an interpolation between the two extreme cases.

\section{Wilson loop correlators and scattering amplitudes}

We saw that an inherent feature of the $q$-$\qb$ scattering amplitude
is its IR divergence. In order to remedy this, and also to show a
context where the finite behaviour of the inelastic amplitudes 
calculated in the previous section appears directly without the
infinite phases,
we are led to consider the scattering of two $q\qb$ 
pairs of transverse size $a$, and impact parameter distance $L$.
This process is interesting to study in itself, since it gives some
information on the scattering amplitudes between colourless states in gauge
theories at strong coupling.

For this setup we have to calculate the correlation function of two
Wilson loops \cite{Nachtr}, where the loops are choosen to be
elongated along the ``time'' 
direction  and have a large but arbitrary temporal length $T$
(the exact analogue for Wilson loops of $T$ considered in
the previous section). However, the cut-off
dependence on  $T$ is expected to be removed  
together with the related IR divergence which was present for the case
of Wilson lines.  

For large positive and negative times the minimal surface will be well
approximated by two seperate copies of the standard minimal surfaces for
each loop separately. When we come to the interaction region, and for $L$
sufficiently small, one can lower the area by forming a ``tube'' joining
the two worldsheets.
Since we want to calculate the normalized correlator $\cor{W_1
W_2}/\cor{W_1}\cor{W_2}$, the contributions of the regions outside the
tube will cancel out (in a first approximation neglecting deformations
near the tube). Therefore we have just to find the area
of the tube, and subtract from it the area of the two independent
worldsheets. It is at this stage that we see that the result does not
depend on the maximal length of the Wilson loops $T$, and hence is IR
finite. The whole contribution to the amplitude will just come from
the area of the tube.

Since we cannot obtain an exact minimal surface for these boundary
conditions, let us perform a variational approximation. Namely we will
consider a family of surfaces forming the tube, parameterized by $T_{tube}$,
which has the interpretation of an ``effective'' time of
interaction. Then we will make a saddle point minimization of the area
as a function of this parameter.

Suppose that the tube linking the two Wilson lines is formed in the
region of the time parameter $t\in (-T_{tube},T_{tube})$. In our
approximation its two
``sides'' are formed by sheets of the helicoid solution (of
area $S(T_{tube}),$ see (\ref{e.ahelic}) for the euclidean case and
(\ref{e.ahelicmin}) for the minkowskian one). The front and back will
be each approximated by strips of area $aL \sqrt{1+\f{T^2_{tube}
\th^2}{L^2}}$ (we assume $a,L \geq R_0$).

The total area corresponding to the two Wilson loops is then given by 
\eq
\label{e.tube}
Area(T_{tube}) = 2L\int_{-T_{tube}}^{T_{tube}}d\tau \sqrt{1+
\ttl}  +2aL \sqrt{1+\f{T^2_{tube} \th^2}{L^2}} -4a
\cdot T_{tube} \ ,
\eqx
where $-2 a T_{tube}$ is the contribution of each individual Wilson loop
to the normalization $1/\cor{W_1}\cor{W_2}$ of the Wilson loop
correlation function. 

Analytically continuing the area formula (\ref{e.tube}) to the
Minkowskian case and using a convenient change of variables, the
Minkowskian area can be put in the following simple form
\eq
\label{e.tubeminkow}
Area(T_{tube}) = \f{2L^2}{\chi} \left\{ \phi+\f{\sin 2\phi}{2}+\rho
\chi \cos\phi -2\rho \sin \phi \right\} \ ,
\eqx
where $\rho\equiv a/L$ and $\sin \phi=i\chi\, T_{tube}/L$ is
the new variational parameter.

In the strong coupling limit ($\al' =1/\sqrt{2g^2_{YM}N} \!\to\! 0$)
the parameter $\phi$ is dynamically determined from the saddle point equation: 
\eq
\label{e.sp}
0=\f{\partial Area(\phi)}{\partial \phi}=
\cos\phi(\cos\phi-\rho)-\f{\rho \chi}{2}\sin\phi
\eqx
It is easy to realize that for large enough energy, there  exists a
solution with $\phi \sim \pm n\pi$. Inserting this solution into the area
(\ref{e.tubeminkow}) we find 
\eq
\label{e.aphi}
Area(\phi)=-\f{2L^2}{\chi} n\pi+2aL (-1)^n
\eqx
where we retain the physical solutions with $n$ positive integer. We
thus find a set of solutions very similar to the inelastic factor
obtained in section {\bf 2}. The modification due to the front-back
contribution $2aL$ is negligible in the Fourier transformed amplitude
for momentum transfer $\sqrt{-t} \gg a/R_0^2$. Also this term is
probably more dependent on the treatment of the front-back parts of
the tube in our approximation.

It is interesting to note that the minimization (\ref{e.sp}) gives
rise in a natural way to a similar set of solutions parameterized by
integers as found from the branch cut arguments in section {\bf 2}.
Each value of $n$ corresponds to a saddle point i.e. a classical
solution. The determination of the weights of each component to the
total scattering amplitude is beyond the reach of the classical
approximation.

For completeness, let us briefly discuss the general saddle point
solution. For lower energies, there are families of solutions also leading to
reggeized behaviour but with distorted trajectories. For
$\chi$ small there exist solutions with $\phi$ imaginary and thus
leading to elastic parts of the amplitude. The study of these
solutions is beyond the scope of the present paper.
For too large impact parameters we may
enter the purely elastic regime found in \cite{ja99} which does not
correspond to connected minimal surfaces (the Gross-Ooguri
transition \cite{Gross}).

As a word of caution (and incentive for further study) 
we note that the saddle point in terms of
$T_{tube}$  is mainly driven to complex values.  
This indicates that a complete treatment and an investigation of the
Gross-Ooguri transition requires a more refined
study of the tube minimal surface.

Let us analyze the properties of the resulting amplitude.
Recalling that charge conjugation acting on one of the $q\qb$ pairs is
equivalent to considering the transformation $\chi \!\to\! \chi-i\pi$,
it is convenient to analyze the components of definite signature
\cite{fr62} with the even and odd contributions given by
\eq
\label{e.pshift}
\tilde T^{\pm}_n(l,s) =\ e^{- n \f{\sqrt{2g_{YM}^2N}}{\chi}
\f{L^2}{R_0^2} }\pm\  
e^{ - n \f{\sqrt{2g_{YM}^2N}}{\chi-i\pi} \f{L^2}{R_0^2}} \ .
\eqx
Note the relative factor of 2 in the exponent in comparison with
(\ref{e.ampim}) due to the two sheet structure of the minimal surface.
  
Using the Fourier transform (\ref{e.four}) we finally get 
\eq
\label{e.fourfinal}
T_n^\pm(s,t)=\f{iR_0^2\ln s}{2n \sqrt{2g^2_{YM}N}}\ s^{\alpha_n(t)} \mp
\f{iR_0^2\ln (-s)}{2n \sqrt{2g^2_{YM}N}}\ (-s)^{\alpha_n(t)} \ ,
\eqx
where
\eq
\label{e.bhregge}
\nonumber\\
\alpha_n(t)= 1+\f{R_0^2}{4n \sqrt{2g^2_{YM}N}}t\  .
\eqx
Let us consider the contribution with $n=1$ which is dominant at large
$L$. It is easy to realize that the amplitude (\ref{e.fourfinal})
corresponds to specific Regge 
singularities in the S-matrix framework, namely  double Regge poles whose 
trajectory is given 
by $\alpha_1(t).$ Indeed, using the usual Mellin transform
$s^{\alpha}\equiv\int \f {s^{j} dj}{2i\pi(j-\alpha)},$ 
it can 
be written in the following equivalent forms:
\eqn
\label{e.fourregge}
\!\!\!\!T_1^{\pm}(s,t)\!\!&\!=\!&\!\!\f{iR_0^2}{ 2\sqrt{2g^2_{YM}N}}\
\f{\partial}{\partial\alpha  
}\left\{s^{\alpha_1(t)}\mp(-s)^{\alpha_1(t)}\right\}\\
\!\!&\!=\!&\!\!\f{R_0^2}{ 2\sqrt{2g^2_{YM}N}} \int_{\cal C} \f {dj}{\pi}\
\f{e^{-i\pi j/2} s^j}{(j-\alpha_1(t))^2}
\left\{\!\!
\begin{array}{c}
i\sin\left(\f{\pi j}{2}\right)\\ 
\cos\left(\f{\pi j}{2}\right)
\end{array}
\!\!\right\}  
 \ ,\nonumber
\eqnx
where the complex contour ${\cal C}$ can be taken around the Regge (di)pole 
trajectory
$\alpha_1(t)$ and the signature factors are either $\sin \pi j/2$ or\\
$-i\cos \pi j/2$ depending on the positive or negative signature.

Let us discuss the contributions $T_n$ to the amplitude with
$n>1$. In the absence of a direct determination of their relative weights,
it is interesting to note that unitarization of Regge
amplitudes in the S matrix framework \cite{vh,bvh} leads to a similar
decomposition where the $T_n$ correspond to Regge pole/cut
singularities. In particular, the overlap matrix formalism \cite{vh},
see (\ref{e.bhv}), leads to a specific model for the relative weights
of the $T_n$'s, if we assume a gaussian distribution $f(l,s)\sim
f_0\exp\left(-\f{\sqrt{2g_{YM}^2N}}{\chi} \f{L^2}{2R_0^2}\right)$
for the inelasticity. In this framework~\cite{vh} unitarity is fulfilled
whenever $0<f_0<1/2$. 
However the derivation of the Wilson line/loop correlation function
does not allow us to give model-independent predictions for these weights in
the total amplitude.

Finally let us comment on the relation of our results on the
trajectory $\alpha(t)$ with the glueball spectrum calculations
\cite{glueballs}.
An extrapolation of the trajectory (\ref{e.bhregge}) to positive $t$ leads to
masses of the form
\eq
M^2 = 4n(J-1) \f{\sqrt{2g^2_{YM}N}}{R_0^2}
\eqx
where $J$ is the spin and $n$ labels the different trajectories.
Because of the appearance of coupling constant dependence it is easy
to see that these states correspond to massive
string states and not to supergravity fields associated with the
glueballs found in \cite{glueballs}. Indeed the latter states have
masses proportional just to $1/R_0^2$ and  spin limited by $J\leq
2$. The appearence of massive string states is not surprising in our
case as we consider an extended string worldsheet between the two
Wilson loops instead of a supergravity field exchange.
The transition between both situations and thus the relation between
both sets of states remains an open problem. 

We should note that our approximations for calculating the
Wilson loop correlator (which is the channel relevant for glueballs)
are rather crude and become problematic at small $t$ (consider the
discussion after (\ref{e.aphi})). Therefore the extrapolation of the
linear trajectory into the glueball regime can easily break
down. Unfortunately the complexity of the minimal surface problem with
the Wilson loop boundary conditions does not allow us to make more
quantitative estimates.

\section{Conclusions and outlook}

Let us give our main conclusions. By computing Wilson line and Wilson
loop correlation functions in the framework of the AdS/CFT
correspondence we show a relation between  minimal surface
problems in $AdS_5$ metrics and reggeization in gauge field theory at
strong coupling. 

For Wilson line correlators, we isolate in certain cases IR finite
inelastic amplitudes coming from the branch cut structure 
of the analytical continuation of helicoid-like surfaces i.e. minimal
surfaces with straight line boundary conditions corresponding to
classical trajectories in Minkowski space. 

We considered three
cases: (i) flat metric approximation of an $AdS$ black hole metric giving
rise to Regge amplitudes with linear trajectories, (ii) an
approximate evaluation for the conformal $AdS_5 \times S^5$ geometry
leading to flat Regge trajectories\footnote{A
remaining IR divergence in the inelastic amplitude is still present in
the absence of confinement.} 
and (iii) evidence for a transition, in a confining theory, from
behaviour of type (i) to (ii)
when the impact parameter decreases below the interpolation scale set by
the horizon radius. In this case, confinement provides a natural IR
cut-off scale.

In a second stage we considered the correlation function of two Wilson
loops elongated along the light cone directions for the confining
geometry. This configuration corresponds to a
high energy scattering amplitude between colourless $q\qb$ states.
We use a variational approximation where the minimal surface is
constructed from two helicoidal sheets. As expected, the obtained
amplitude is free from IR divergences and gives rise to reggeization
with a linear trajectory with unit intercept. For high energies the
amplitude is imaginary and thus mainly reflects the inelasticity of
the process.

These results call for some comments. 

We note that the structure of our resulting amplitudes for the
confining case 
(in particular the $n$, $\chi$ and $L$ dependence) matches the 
calculations of the
imaginary part of flat space D-brane scattering amplitudes \cite{ba94}
and some specific Wilson loop correlators\footnote{We have extracted the
imaginary part  from the formulae in Ref. \cite{ch99}, along the
lines of \cite{ba94}.} \cite{ch99}, when the
``effective'' string length $\sqrt{\al'}$ is taken to be set by the
horizon radius in our case. It is interesting to note that the
imaginary part in those calculations is generated from the
singularities of the string amplitudes which are an infinite set of
poles. The slopes of the trajectories are the same as in our case
(\ref{e.bhregge}), while the intercepts are different. However the
geometrical configurations in \cite{ch99} is quite different from the
one we considered in section {\bf 4}. Even in the flat space approximation
it would be useful to have a direct string calculation of the tube
configuration.

Beyond the flat space approximation, we want to 
emphasize the interest of solving exactly the well 
defined mathematical problem of finding the generalization of the helicoid 
for various AdS metrics, i.e. the minimal surface
spanned between infinite lines forming an angle $\th$ at the boundary. 
Another goal is to go beyond the classical approximation in order to
derive the $n$-dependent weights to the scattering amplitudes.

Indeed the generalization of the helicoidal geometry in AdS space
seems to be a building block for high energy scattering amplitudes in
gauge theories at strong coupling.

\subsubsection*{Acknowledgements}

RJ was partially supported by KBN grants 2P03B00814,
2P03B08614.  We thank T. Garel and B. Giraud for useful remarks.


\begin{thebibliography}{99}

\bibitem{ma98} \rr{J. Maldacena}{Adv. Theor. Math. Phys.}{2}{(1998)
231};\\ 
\rr{S.S. Gubser, I.R. Klebanov and
A.M. Polyakov}{Phys. Lett.}{B428}{(1998) 105};\\
\rr{E. Witten}{Adv. Theor. Math. Phys.}{2}{(1998) 253}.

\bibitem{ma99} \rr{O. Aharony, S.S. Gubser, J. Maldacena, H. Ooguri
and Y. Oz}{Large $N$ field theories, String Theory and
Gravity,}{}{hep-th/9905111}.

\bibitem{fr62}\rr{M. Froissart and R. Omnes}{Mandelstam Theory and Regge 
Poles,}{}{Frontiers in Physics, Benjamin ed. 1963}.
      
\bibitem{fr61}
\rr{M.Froissart}{Phys.Rev.}{123}{(1961) 1053}.
      

\bibitem{ja99}
\rr{R.A. Janik and R. Peschanski}{Nucl. Phys.}{B565}{(2000) 193};
\rr{R.A. Janik}{Gauge Theory Scattering from the
AdS/CFT correspondence, Cargese summer school 1999,}{}{hep-th/9909124}. 

\bibitem{Nacht} \rr{O. Nachtmann}{Ann. Phys.}{209}{(1991) 436}.
\bibitem{VV} \rr{H. Verlinde and E. Verlinde}{QCD at High Energies and
Two-Dimensional Field Theory,}{}{hep-th/9302104}.
\bibitem{kor} \rr{G.P. Korchemsky}{Phys. Lett.}{B325}{(1994) 459}.
       

\bibitem{Nachtr} \rr{O. Nachtmann}{High
Energy Collisions and Nonperturbative QCD,}{}{hep-ph/9609365} (see
e.g. eq. (3.87) for colourless state scattering).
\bibitem{Wilson} \rr{J. Maldacena}{Phys. Rev. Lett.}{80}{(1998) 4859};\\
\rr{S.-J. Rey and J. Yee}{Macroscopic strings as heavy quarks in large
$N$ gauge theory and anti-de Sitter supergravity,}{}{hep-th/9803001}.


\bibitem{wi98}\rr{E. Witten}{Adv. Theor. Math. Phys.}{2}{(1998) 505}.

\bibitem{rey} \rr{S.-J. Rey, S. Theisen and
J.-T. Yee}{Nucl. Phys.}{B527}{(1998) 171} 

\bibitem{brand} \rr{A. Brandhuber, N. Itzhaki, J. Sonnenschein and
S. Yankielowicz}{Phys. Lett.}{B434}{(1998) 36} 

\bibitem{Gross} \rr{D.J. Gross and H. Ooguri}{Phys. Rev.}{D58}{(1998) 106002}.

\bibitem{estimates} \rr{Y. Kinar, E. Schreiber and
J. Sonnenschein}{Nucl. Phys.}{B566}{(2000) 103}; see also section 11
in \cite{so99}.


\bibitem{Zahed} \rr{M. Rho, S.-J. Sin and I. Zahed}
{Phys. Lett.}{B466}{(1999) 199}.

\bibitem{Megg} \rr{E. Meggiolaro}{Z. Phys.}{C76}{(1997) 523}, 
\rr{}{Eur. Phys. J.}{C4}{(1998) 101},
\rr{}{Phys. Rev.}{D53}{(1996) 3835}.

\bibitem{de99} \rr{H. Ooguri, H. Robins and
J. Tannenhauser}{Phys. Lett.}{B437}{(1998) 77}

\bibitem{ks} \rr{A. Kehigas and K. Sfetsos}{Phys. Lett.}{B456}{(1999) 22}

\bibitem{carlo} \rr{C. Angelantonj and A. Armoni}{RG Flow, Wilson
Loops and the Dilaton Tadpole,}{}{hep-th/0003050}

\bibitem{so99}\rr{Y. Kinar, E. Schreiber, J. Sonnenschein and
N. Weiss}{Quantum fluctuations of Wilson loops from string
models,}{}{hep-th/9911123}. 

\bibitem{fomenko} \rr{A.T. Fomenko}{The Plateau Problem Part 1 and
2,}{}{GordonBreach 1989}

\bibitem{bo99}\rr{A. Boudaoud, P. Patr\`\i cio, and M. Ben Amar}{Phys. Rev. 
Lett.}{83}{(1999) 3836}. 

\bibitem{ba94} \rr{C. Bachas}{Phys. Lett.}{B374}{(1996) 37}.
\bibitem{Messiah} E.g. see \rr{A. Messiah}{Quantum Mechanics, vol
I and II}{}{North Holland 1961}.

\bibitem{qedeik} \rr{H. Cheng and
T.T. Wu}{Phys. Rev. Lett.}{22}{(1969) 666};\\
\rr{H. Abarbanel and C. Itzykson}{Phys. Rev. Lett.}{23}{(1969) 53}.
\bibitem{bvh} \rr{A. Bia{\l}as and L. van Hove}{Nuovo Cim.}{38}{(1965) 1385}

\bibitem{vh} \rr{L. van Hove}{Rev. Mod. Phys.}{36}{(1964) 655}

\bibitem{korch} \rr{I.A. Korchemskaya}{Nucl. Phys.}{B490}{(1997) 306}

\bibitem{neww} \rr{N. Drukker, D.J. Gross and H. Ooguri}{Phys.Rev.,}{D60}{(1999) 
125006};\rr{H. Ooguri}{Prog.Theor.Phys.Suppl.,}{134}{(1999) 153}.

\bibitem{BFKL} \rr{L.N. Lipatov}{Sov. J. Nucl. Phys.}{23}{(1976) 642};
\rr{V.S. Fadin, E.A. Kuraev and L.N. Lipatov}{Phys. Lett.}{B60}{(1975)
50}; \rr{E.A. Kuraev, L.N. Lipatov and V.S. Fadin}{Sov. Phys. JETP}{44}{(1976)
45, {\bf 45} (1977) 199}; \rr{I.I. Balitsky and
L.N. Lipatov}{Sov. J. Nucl. Phys.}{28}{(1978) 822}.\\
For a discussion in impact parameter space see
\rr{G.P.Salam}{Quarkonium Scattering at High Energies,}{}{PhD thesis,
University of Cambridge 1996}.

\bibitem{bern} \rr{Z. Bern, J. Rozowsky and B. Yan}{Two loop
\mbox{${\cal N}=4$} Supersymmetric Amplitudes and
QCD,}{}{hep-ph/9706392, talk given at \mbox{DIS 97}}.

\bibitem{glueballs} \rr{C. Csaki, H. Ooguri, Y. Oz and
J. Terning}{JHEP}{9901}{(1999) 017};
\rr{R. de Mello Koch, A. Jevicky, M. Mihailescu and
J.P. Nunes}{Phys. Rev.}{D58}{(1998) 105009}.\\       
For a recent extension to spin two glueballs see \rr{R.C. Brower,
S.D. Mathur and C.-I Tan}{Glueball Spectrum for QCD from AdS
Supergravity Duality,}{}{hep-th/0003115};
\rr{N.R. Constable and R.C. Myers}{JHEP}{9910}{(1999) 037}.

\bibitem{ch99} \rr{S. Chaudhuri, Y. Chen and E. Novak}{Pair
Correlation Function of Wilson Loops}{}{hep-th/9910183};
\rr{S. Chaudhuri and E. Novak}{Supersymmetric Pair
Correlation Function of Wilson Loops}{}{hep-th/0002046}.

\end{thebibliography}
\end{document}